\documentclass[prl,aps,twocolumn,groupedaddress,showpacs]{revtex4}

\begin{document}

\input psfig.sty

\title{Head-on/Near Head-on Collisions of Neutron Stars With a Realistic EOS}

\author{Edwin Evans${}^{(1)}$, A. Gopakumar${}^{(1,2)}$, Philip Gressman${}^{(1,3)}$, Sai Iyer${}^{(1)}$, Mark Miller${}^{(1,4)}$, Wai-Mo Suen${}^{(1,5)}$, and Hui-Min Zhang${}^{(1)}$}

\address{${}^{(1)}$McDonnell Center for the Space Sciences,
Department of Physics,
Washington University, St. Louis, Missouri 63130}

\address{${}^{(2)}$Physics Department,
University of Guelph, Canada}
\address{${}^{(3)}$Mathematics Department,
Princeton University,
Princeton, NJ 08544}
\address{${}^{(4)}$ 238-332 Jet Propulsion Laboratory,
4800 Oak Grove Drive,
Pasadena, CA 91109}
\address{${}^{(5)}$Physics Department,
Chinese University of Hong Kong,
Hong Kong}

\date{\today}

\begin{abstract}

It has been conjectured that in head-on collisions of neutron stars (NSs),
the merged object would not collapse promptly even if the total mass
is higher than the maximum stable mass of a cold NS.  In this paper,
we show that the reverse is true: even if the total mass is {\it less}
than the maximum stable mass, the merged object can collapse promptly.
We demonstrate this for the case of NSs with a realistic equation of
state (the Lattimer-Swesty EOS) in head-on {\it and} near head-on
collisions.  We propose a ``Prompt Collapse Conjecture'' for a generic
NS EOS for head on and near head-on collisions.

\end{abstract}

\pacs{04.25.Dm,04.30.+x,97.60.Jd,97.60.Lf}

\maketitle



\paragraph*{\bf Introduction.}
\label{introduction}

Shapiro \cite{Shapiro98a} has conjectured that two neutron stars (NSs)
falling from infinity and colliding head-on
would not, independent of their masses, promptly collapse to a black hole.  The
basic argument for the conjecture is that, until there is significant 
neutrino cooling, the thermal pressure
generated by shock heating is always enough to support the merged
object.   Since neutrino cooling operates on
time scales of seconds instead of the NS dynamical time scale of
milliseconds, this could have significant implications on
gravitational wave and neutrino emissions.  The results obtained 
in~\cite{Shapiro98a}
were based on a polytropic equation of state (EOS) $P =
K \rho^\Gamma$ (with $K$ a function of the entropy), and it was suggested that the
result was true for a general EOS.

We have shown in~\cite{Miller01a} that one implicit assumption used in
the derivation in \cite{Shapiro98a}, namely that the collision can be
approximated by a quasi-equilibrium process, is not valid.  We carried
out simulations of the head-on collision of neutron stars described by
a polytropic EOS, as in the conjecture.  For two 1.4 $M_{\odot}$ NSs
(with a polytropic index of $\Gamma=2$, initial polytropic coefficient
$K$ of $1.16 \times 10^5 \; {\rm cm}^5/{\rm g} \; {\rm s}^2$ as in a
typical NS model), we showed that the merged object collapsed
promptly.  The shock front generated in the collision does not even
have time to propagate to the outer part of the merged object before
it is engulfed in an apparent horizon, let alone produce enough
thermal pressure to support the merged object as envisioned in the
quasi-equilibrium argument of \cite{Shapiro98a}.  In our study in
\cite{Miller01a}, the merged object had a mass well above the critical
mass (the maximum mass that the EOS with the polytropic constants
$\Gamma$ and $K$ given above can support).  We pointed out in
\cite{Miller01a,Miller99c} that the prompt collapse is due to the
dynamical compression of the collision that is absent in the
quasi-equilibrium analysis in \cite{Shapiro98a,Shapiro99a}.

This brings up three further questions: 1. What if the
mass of the merged object is {\it less} than the critical mass?  Will
the dynamical compression in the collision process be strong enough to
initiate a collapse?  2. What if we use a realistic EOS instead of a
polytropic EOS?  3. What if we break the exact axisymmetry?  In this
paper we answer these three questions with one set of
numerical simulations.  The simulations are based on the GR-Astro code
(formerly called GR-3D) constructed in the NASA Neutron Star Grand
Challenge project \cite{NASA} and the NSF Astrophysics Simulation
Collaboratory (ASC) project \cite{ASC}.  For the construction of the
code and the classes of validation tests we have carried out for it,
see \cite{Font98b,Font98f,Font01b,Suen99a}.  The GR-Astro code solves the
coupled set of the Einstein equations and the general relativistic
hydrodynamic equations with a realistic EOS.  It will be released to
the community through the ASC portal \cite{ASC} upon completion of the project.

In this paper, we report on simulations of NSs constructed with the
Lattimer-Swesty\cite{Lattimer91} (LS) EOS, which has been used in
various neutron star studies (all existing simulations based on the LS
EOS that we are aware of are based on Newtonian gravity and hence
cannot answer questions of collapse).  To the best of our knowledge, our simulations represent the first set of general relativistic 3D simulations based on a realistic EOS (for polytropic EOS simulations, see \cite{Shibata99d}).  In this study, we use
neutron stars of rest (baryonic) mass 1.6 $M_{\odot}$ (corresponding to
an ADM mass of $1.4 \; M_{\odot}$ in isolation), with a radius of
$13.8 \;{\rm km}$ (proper distance).  The merged object has a rest mass
of 3.2 $M_{\odot}$ which is considerably {\it lower} than the critical
mass of 3.67 $M_{\odot}$ in the LS EOS with our choice of
parameters.
Nevertheless, we find that the merged object will
promptly collapse to a black hole within a dynamical timescale.  An
apparent horizon is found engulfing the shock wave at $0.15 \;{\rm ms}$
after the two stars have touched, with time measured at infinity,
i.e. at the edge of the computational grid, which is $47 \;{\rm km}$ away
from the collision center.  The prompt collapse is verified for both
the head-on collision case and an off-axis collision (the axisymmetry
is broken by an impact parameter of 1/2 stellar radius).  This study
demonstrates that dynamical effects are strong enough to cause a
prompt collapse even when the total mass is {\it below} the 
single-star critical mass, and that this result is
{\it not} a consequence of exact axisymmetry.  We hence post the
following ``Prompt Collapse Conjecture'': 
For head-on and near head-on collisions of
neutron stars described by a generic equation of state and infalling from
rest at infinity, there exists a window in the rest mass of the merged
object, {\it below} the critical single-star rest mass, where prompt
collapse to a black hole can occur.
The claim of the ``near head-on'' part means that the prompt collapse is stable with respect to small perturbations of the initial velocity.
\paragraph*{\bf The Setup.}
\label{setup}

In this paper we use our implementation of the LS EOS as described in
\cite{NASA_EOS}. We use the LS EOS in a tabular form with rest mass
density and specific energy density as the two independent
thermodynamic variables, which are evolved using the general
relativistic hydrodynamic (GR-Hydro) equations (the lepton to baryon
ratio is set to a constant (0.1) in all simulations in this paper). As
in the Newtonian LS EOS simulations of \cite{Ruffert96b,Ruffert97a},
we set the initial specific energy density of the NSs at a
relatively large value of $0.9 \;{\rm Mev}$.  The ADM mass and rest
mass as a function of central rest mass density for a single
static NS is given in Fig. 1.
We see that the critical rest mass of an LS EOS star is at 3.67
$M_{\odot}$.  In our simulation we use NSs with a rest mass of 1.6
$M_{\odot}$ (marked by a X in Fig.~1).  The merged object is hence
guaranteed to have a rest mass below the critical single-star mass.  In Fig.~1
we have also plotted the corresponding curves for the case of a
polytropic EOS with $\Gamma=2$ and $K =1.16 \times 10^5 \; 
{\rm cm}^5/{\rm g} \; {\rm s}^2$ for comparison.

\begin{figure}
\begin{center}
\vspace{0.5cm}
\psfig{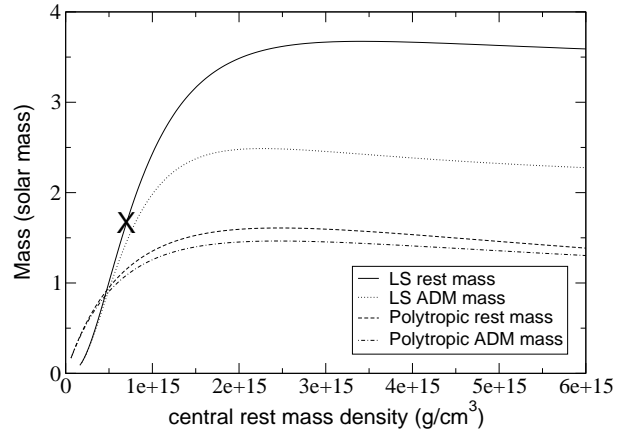}
\end{center}
\caption{The ADM mass and
rest mass vs. central rest mass density of
a static NS with the LS EOS.  The corresponding curves for a polytropic EOS 
with $\Gamma=2$, $K =1.16 \times 10^5 \;{\rm cm}^5/{\rm g} \; {\rm s}^2$
are plotted for comparison. }
\end{figure}

In the head-on collision case, we put the two stars at a proper
distance of $d=42\;{\rm km}$ apart (about $3R$ separation, $R$ = radius
of star) along the z-axis, and boost them toward one another at the
speed (as measured at infinity) of $\sqrt {GM/d}$ (the Newtonian
infall velocity).  The metric and extrinsic curvature of the two
boosted stars are obtained by (i) adding the off-diagonal components
of the metric, (ii) adding the diagonal components of the metric and
subtracting 1, (iii) adding the components of the extrinsic curvature.
The resulting matter distribution, momentum distributions,
conformal part of the metric, and transverse traceless part of the
extrinsic curvature are used as input to York's procedure
\cite{York79} for determining the initial data.  The initial data
satisfies the complete set of Hamiltonian and momentum constraints to
high accuracy (terms in the constraints cancel to $10 ^ {-6}$), {\it
and} represents two NSs in head-on collision.

The initial data is then numerically evolved by solving the coupled 
Einstein \ GR-Hydro evolution equations with numerical methods
described in \cite{Font98b}.  The simulations reported here use the
``$1+\log$'' slicing \cite{Anninos94c}.  The simulations have been carried
out with resolutions ranging from $\Delta x = 0.74
\;{\rm km}$ to $0.3 \;{\rm km}$ (28 to 70 grid points across each NS) for
convergence and accuracy analysis. We find that the constraint
violations rise linearly throughout the evolution, and converge to
zero with increased numerical resolution. The
total rest mass of the system is conserved to better than 0.2\% throughout
the simulations.

\paragraph*{\bf The Results.}
\label{result}

In Fig.~2a, we show the collapse of the lapse along the $z$ axis from
$t=0 \;{\rm ms}$ to $t=0.37 \;{\rm ms}$ at intervals of $0.0926 \;{\rm
ms}$. (With the reflection symmetry across the $z=0$ plane and the
axisymmetry of the head-on collision, we only need to evolve the first
octant.)  By the time $t=0.37 \;{\rm ms}$, 
the lapse has collapsed significantly.
Fig.~2b shows the corresponding evolution of the $zz$ component of the
metric function.  The ``grid stretching'' peak, characteristic of a
black hole evolved in a singularity avoiding slicing, is apparent.

\begin{figure}
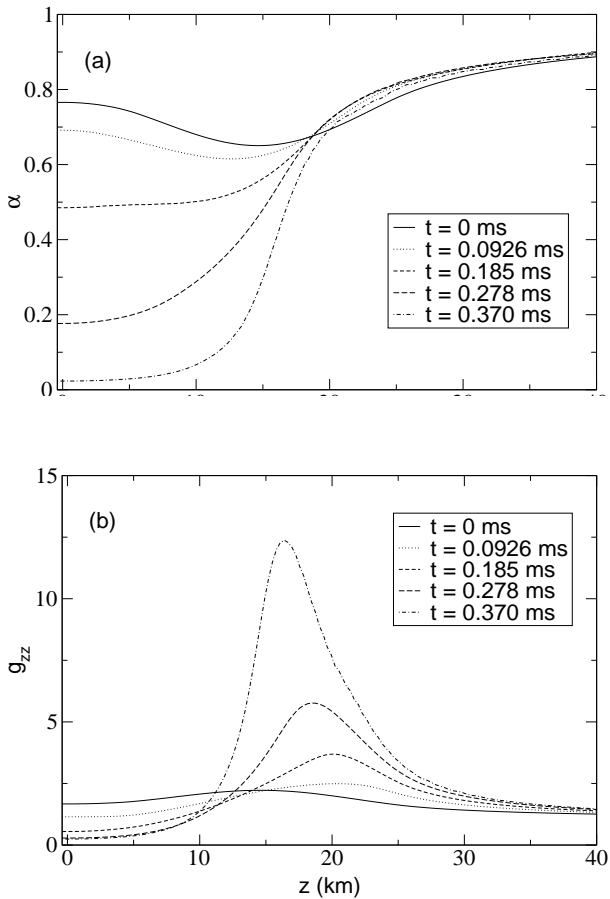

\begin{center}
\vspace{0.25cm}
\psfig{figure=headon_lapse.eps,width=8.0cm}
\vspace{0.25cm}
\psfig{figure=headon_gzz.eps,width=8.0cm}
\end{center}
\vspace{-0.5cm}
\caption{
The lapse $\alpha$ and metric component $g_{zz}$ is displayed
      along the $z$-axis at
      various times for the head-on collision of 2 NSs using the LS EOS.  
      This simulation used
      $163^3$ grid points, with $\Delta x = 0.3 \; km$.  
}
\end{figure}

Fig.~3 shows contour lines in the $y=0$ plane of the log of the
gradient of the rest mass density
$\log\left(\sqrt{\nabla^i(\rho)\nabla_i(\rho)}\right)$ at time $t
=0.37 \;{\rm ms}$.  Sharp changes in rest mass density (where contour
lines bunch up) indicate shocks.  We see that the shock front is at
$7\;{\rm km}$ in the x direction, and $10\;{\rm km}$ along the $z$ direction
(the collision axis) and has not yet reached the back end of star (at
$14.5\;{\rm km}$).  At this point the shock front is still moving outward in
coordinate space, although it is completely engulfed by the apparent
horizon, as seen in Fig.~4 below.

\begin{figure}
\begin{center}
\vspace{-1.5cm}
\psfig{figure=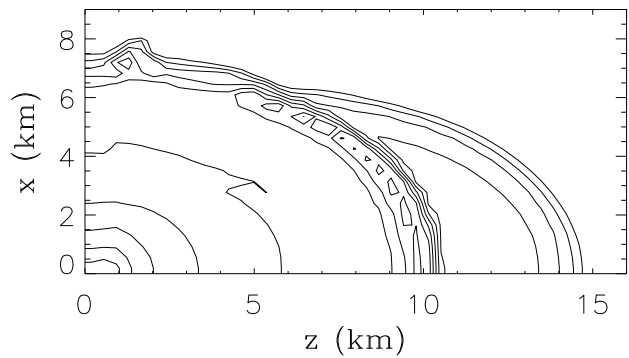,width=9cm}
\end{center}
\vspace{-1.0cm}
\caption{ Equally spaced contour lines of the log of the 
gradient of the rest mass
density $\log\left(\sqrt{\nabla^i(\rho)\nabla_i(\rho)}\right)$,
showing the shock front at $t =0.37 \; ms $.  }
\label{fig:shock}
\end{figure}

\begin{figure}
\begin{center}
\vspace{0.5cm}
\psfig{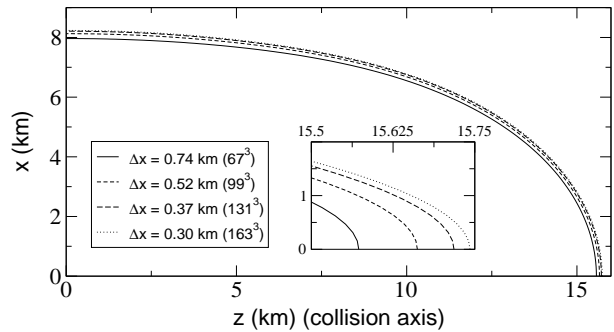}
\end{center}
\vspace{-0.25cm}
\caption{ The position of the AH in the $x-z$ plane 
at different resolutions, all at $t = 0.37 \; ms$.
}
\label{fig:horizon}
\end{figure}

In Fig.~4, we show the intersection of the apparent horizon (AH) with
the $x-z$ plane.  To confirm the location of the AH, convergence
tests, both in terms of resolution and in terms of location of the
computational boundary, have been carried out.  The solid, short
dashed, long dashed and dotted lines correspond to the AH locations at
resolutions of $\Delta x = 0.74, 0.52, 0.37 $ and $0.30 \;{\rm km}$.
The inset is a close-up view near the $z$-axis.  We see that the
location of the AH converges in a first order manner.  We emphasize
the importance of demonstrating the convergence of the final results
in a numerical simulation.  Such a validation of the final result is
computationally expensive but nevertheless indispensable, as a short
term convergence test of the numerical code does {\it not} guarantee
the convergence of the final result at late time.  In Fig.~5 we show
the hamiltonian constraint converging linearly with respect to
resolution for the whole duration of the numerical evolution.

\begin{figure}
\begin{center}
\vspace{0.5cm}
\psfig{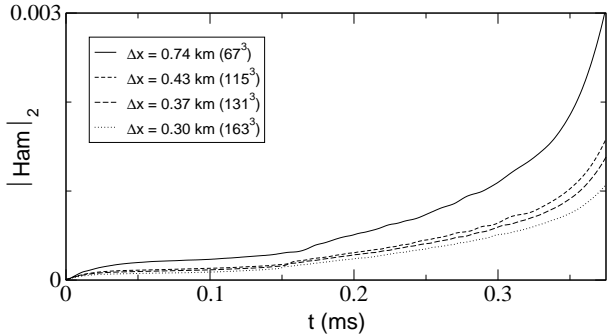}
\end{center}
\vspace{-0.25cm}
\caption{ The L2 norm of the hamiltonian 
constraint converging 
linearly with respect to resolution for the whole duration of the
numerical evolution. 
}
\label{fig:ham}
\end{figure}

\paragraph*{\bf Analysis of the Results.}
\label{analysis}

The above results demonstrate the phenomenon of a ``sub-critical mass
collapse'': a prompt collapse of the merged object, despite the fact
that its total rest mass is {\it below} the critical rest mass of a
single NS.
 
One might question whether this result depends sensitively on the
choice of the initial velocity.  As the star was only at about $3R$
separation at the start of the simulation the infall velocity
determined by the Newtonian formula is not accurate. We have verified
that increasing or decreasing the initial velocity by 10\% does not
change the qualitative feature of the collision, and in particular the
prompt collapse result.  Likewise, our initial configuration while
satisfying the constraints may not have exactly the same gravitational
wave content and distortion of the NSs as one that represents free
fall from infinity.  We do not expect that to have a significant
influence on the result of prompt collapse (although it will affect
the gravitational waveforms).  To verify this point we have also
carried out simulations using slightly distorted stars and slightly
modified extrinsic curvature for the construction of the initial data.
We confirmed that the prompt collapse result is not affected.

One might also wonder whether this prompt collapse result is a
consequence of the exact symmetry used.  To determine this point we
carried out simulations with the two initial NS moving in the $z$
direction but offset in the $x$ direction by half a radius (an impact
parameter of $6.9\;{\rm km}$).  With this setup, the computation
expense is significantly higher, as we can no longer evolve just an
octant as in the axisymmetric case above.  We found that the inclusion
of a small impact parameter does not change the qualitative features
of the collision, including the occurrence of prompt collapse.  In
Fig.~6, we show the apparent horizon found again at $t=0.37\;{\rm
ms}$.  We see that the AH is tilted with respect to the $z$ axis, but
is very similar in shape and size to the head-on case.

\begin{figure}
\begin{center}
\vspace{1cm}
\psfig{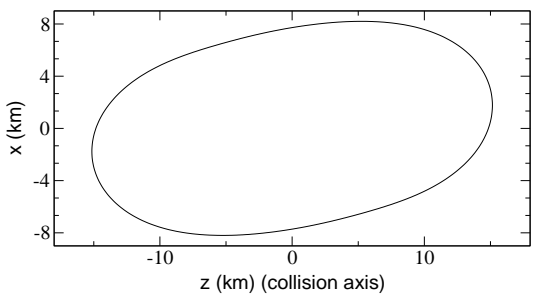}
\end{center}
\caption{ The position of the AH at $t = 0.37 \; ms$ for a near head-on 
collision using a spatial resolution of $\Delta x =  0.37 \; km$.  
}
\label{fig:horizon2}
\end{figure}

In all the above cases, we have prompt collapses.  We have confirmed
that at a low enough mass, the merged object will not collapse but will
instead merge, bounce, oscillate and form a stable NS.  However, the
determination of the dividing line between a final state 
black hole or a final state NS would require 
extremely high resolution whose computational expense is beyond what 
is available to our group.  

Finally, one may also ask whether the sub-critical mass collapse occurs
in the head-on collision of NSs described by a polytropic EOS.
Preliminary studies with our 3D code show that, in fact, the
dividing line in the rest mass of the final merged object between prompt
collapse and non-prompt collapse is
quite close to the critical rest mass of a single NS for the $\Gamma = 2$
polytrope.  We are currently working on an axisymmetric version of
the code which we will use to not only explore this question, but
to also explore the possible existence of type-1 critical
phenomena at the interface dividing the prompt collapse and 
non-prompt collapse cases.  

\paragraph*{\bf Conclusions.}
\label{conclusion}

In this paper we report on a sub-critical mass collapse phenomenon in
the head-on/near head-on collisions of NSs with a realistic EOS.  We
propose a ``Prompt Collapse Conjecture'': 
For head-on and near-head-on collisions of
neutron stars described by a generic equation of state and infalling from
rest at infinity, there exists a window in the rest mass of the merged
object, {\it below} the critical single-star rest mass, where prompt
collapse to a black hole can occur.
This is the opposite of the Shapiro
conjecture~\cite{Shapiro98a}, which predicted no prompt 
collapse for all NS masses,
including those above the single-star critical mass.

It has been argued in \cite{Katz96a} that head-on/near head-on
collisions of NSs could have a significant event rate, and could be a
candidate for a sub-class of short gamma-ray bursts.  The results of prompt
collapse reported in this paper could have implications on the
observation of such processes, with the prompt formation of the
horizon cutting off the causal connection of the shock heated matter
from outside observers.  Note that we are not claiming a prompt
collapse in inspiral coalescence of NSs.

\paragraph*{\bf Acknowledgements.}
\label{ack}

We thank Luc Blanchet, K. Thorne, and C. Will for useful discussions,
and Lap-Ming Lin for comments on the manuscript.  The simulations in
this paper have made use of code components developed by several
authors: BAM (multigrid solver) by B. Br{\" u}gman; AH-FINDER
(apparent horizon finder) by M. Alcubierre; CONF-ADM (evolution of the
Einstein field equations), MAHC (evolution for the GRHydro equations),
and IVP (conformal constraint solver) by M. Miller; PRIM-SOL (solver
for the hydrodynamical primitive variables) by P. Gressman, ELS
(LS-EOS tabular treatment) by E. Evans, and the CACTUS Computational
Toolkit by T. Goodale {\it et al}.

\noindent Support for this research has been
provided by the NSF KDI Astrophysics Simulation Collaboratory (ASC)
project (Phy 99-79985), NASA Neutron Star Grand Challenge
Project(NCCS-153), the NSF NRAC Project Computational General
Relativistic Astrophysics (93S025), and the NASA AMES NAS.

\vspace{-0.5cm}
\bibliographystyle{prsty}

\begin{thebibliography}{1}

\bibitem{Shapiro98a}
S. Shapiro, Phys. Rev. D {\bf 58}, 103002 (1998).

\bibitem{Miller01a}
M. Miller, W.-M. Suen and M. Tobias, Phys. Rev. D {\bf 63}, 121501(R) (2001)

\bibitem{Miller99c}
M. Miller, W.-M. Suen and M. Tobias, gr-qc/9910022 (1999)

\bibitem{Shapiro99a}
S. Shapiro, gr-qc/9909059 (1999)

\bibitem{NASA}
http://wugrav.wustl.edu/Relativ/nsgc.html

\bibitem{ASC}
http://wugrav.wustl.edu/ASC/

\bibitem{Font98b}
J.~A. Font, M. Miller, W.~-M. Suen and M. Tobias,
Phys. Rev. D {\bf 61}, 044011 (2000)

\bibitem{Font98f}
J.~A. Font, M. Miller, W.~-M. Suen and M. Tobias,
Phys. Rev. D Repository, EPAPS: E-PRVDAQ-61-029004 (2000)

\bibitem{Font01b}
J. A. Font, T. Goodale, S. Iyer, M. Miller, L. Rezzolla, E. Seidel,
N. Stergioulas, W-M. Suen and M. Tobias, Phys. Rev. {\bf D 65}, 084024 (2002)

\bibitem{Suen99a}
W.-M. Suen, Prog. Theor. Phys. Suppl. {\bf 136}, 251 (1999)

\bibitem{Lattimer91}
J. M. Lattimer and D. F. Swesty, Nucl. Phys. A, {\bf 535}, 331 (1991)

\bibitem{Shibata99d}
M. Shibata and K. Uryu, Phys. Rev. D, {\bf 61}, 064001 (2000)

\bibitem{NASA_EOS}
http://wugrav.wustl.edu/Codes/GR3D/GR3D\_EOS.html

\bibitem{Ruffert96b}
M. Ruffert, H.-Th. Janka and G. Sch{\"a}fer, Astron. Astrophys.
{\bf 311}, 532 (1996)

\bibitem{Ruffert97a}
M. Ruffert, H.-Th. Janka, K. Takahashi and G. Sch{\"a}fer,
Astron. Astrophys. {\bf 319}, 122 (1997)

\bibitem{York79}
J. York,  in {\em Sources of Gravitational Radiation}, edited by L. Smarr
  (Cambridge University Press, Cambridge, 1979).

\bibitem{Anninos94c}
P. Anninos, K. Camarda, J. Mass{\'o}, E. Seidel, W.-M. Suen and J. Towns, 
Phys. Rev. D {\bf 52}, 2059 (1995)

\bibitem{Katz96a}
J. I. Katz and L. M. Canel, Ap. J. {\bf 471}, 915 (1996)

\end{thebibliography}

\end{document}